 \documentclass[journal, comsoc]{IEEEtran}

\usepackage{gensymb}
\usepackage{cite}
\usepackage{amsmath,amssymb,amsfonts}
\usepackage{graphicx}
\usepackage{xcolor}
\usepackage{kotex}  
\usepackage{caption}
\usepackage{subcaption}
\usepackage{caption}
\usepackage{subcaption}

\usepackage{stfloats}
\usepackage{float}
\usepackage{wrapfig}
\usepackage{xcolor}
\usepackage{colortbl} 
\usepackage{color,soul}
\usepackage{url}
\usepackage{verbatim}
\usepackage{graphicx}
\usepackage{cite}
\usepackage{epstopdf}

\usepackage[acronym,toc,shortcuts]{glossaries}
\usepackage{multirow}
\usepackage{array}
\usepackage{lipsum}

\newacronym{BS}{BS}{Base Stations}
\newacronym{NTN}{NTN}{Non-Terrestrial Networks}
\newacronym{TN}{TN}{Terrestrial Networks}
\newacronym{AmBC}{AmBC}{Ambient Backscatter Communication}
\newacronym{NOMA}{NOMA}{Non-Orthogonal Multiple Access}
\newacronym{PD-NOMA}{PD-NOMA}{Power Domain Non-Orthogonal Multiple Access}
\newacronym{UEs}{UEs}{User Equipments}
\newacronym{UE}{UE}{User Equipment}
\newacronym{UAVs}{UAVs}{Uncrewed Aerial Vehicles}
\newacronym{SDG}{SDG}{Sustainable Development Goals}
\newacronym{UN}{UN}{United Nations}
\newacronym{HAPS}{HAPS}{High Altitude Platform Stations}
\newacronym{AI}{AI}{Artificial Intelligence}
\newacronym{EE}{EE}{Energy Efficiency}
\newacronym{3GPP}{3GPP}{3$^{\text{rd}}$ Generation Partnership Project}
\newacronym{SA1}{SA1}{Service and System Aspects 1}
\newacronym{WG}{WG}{Working Group}
\newacronym{PLMN}{PLMN}{Public Land Mobile Networks}
\newacronym{MNO}{MNO}{Mobile Network Operators}
\newacronym{LEO}{LEO}{Low Earth Orbit}
\newacronym{KPI}{KPI}{Key Performance Indicators}
\newacronym{GEO}{GEO}{Geostationary Equatorial Orbit}
\newacronym{MEO}{MEO}{Medium-Earth Orbit}
\newacronym{CapEx}{CapEx}{Capital Expenditures}
\newacronym{IMS}{IMS}{IP Multimedia Subsystem}
\newacronym{SA}{SA}{System and Architecture}
\newacronym{RIS}{RIS}{Reconfigurable Intelligent Surfaces}
\newacronym{RAN}{RAN}{Radio Access Network}
\newacronym{NGSO}{NGSO}{Non-Geostationary Satellite Orbit}
\newacronym{UPF}{UPF}{User Plane Function}
\newacronym{ISL}{ISL}{Inter-Switch Link}
\newacronym{QoS}{QoS}{Quality-of-Service}
\newacronym{LAN}{LAN}{Local Area Network}
\newacronym{WLAN}{WLAN}{Wide Local Area Network}
\newacronym{ISAC}{ISAC}{Integrated Sensing and Communication}
\newacronym{SNO}{SNO}{Satellite Network Operators}
\newacronym{MENA}{MENA}{Middle East and North Africa}
\newacronym{LTE}{LTE}{Long-Term Evolution}
\newacronym{NAS}{NAS}{Non Access Stratum}
\newacronym{Wi-Fi}{Wi-Fi}{Wireless Fidelity}
\newacronym{FFR}{FFR}{Full Frequency Reuse}
\newacronym{ZF}{ZF}{Zero Forcing}
\newacronym{MIMO}{MIMO}{Multiple-Input-Multiple-Output}
\newacronym{MMSE}{MMSE}{Minimum Mean Square Error}
\newacronym{SSB}{SSB}{Synchronisation Signal Blocks}
\newacronym{CSI}{CSI}{Channel State Information}
\newacronym{CQI}{CQI}{Channel Quality Indicator}
\newacronym{RS}{RS}{Reference Signal}
\newacronym{GNSS}{GNSS}{Global Navigation Satellite System}
\newacronym{AMF}{AMF}{Access and Mobility Management Function}
\newacronym{LMF}{LMF}{Location Management Function}
\newacronym{5GC}{5GC}{5G Core Network}
\newacronym{C-JT}{C-JT}{Coherent Joint Transmission}
\newacronym{NC-JT}{NC-JT}{Non-Coherent JT}
\newacronym{MC}{MC}{Multi-Connectivity}
\newacronym{RRM}{RRM}{Radio Resource Management}
\newacronym{ML}{ML}{Machine Learning}
\newacronym{RU}{RU}{Radio Unit}
\newacronym{CU}{CU}{Control Unit}
\newacronym{DU}{DU}{Distributed Unit}
\newacronym{RIC}{RIC}{RAN Intelligent Controller}
\newacronym{CN}{CN}{Core Network}
\newacronym{EIRP}{EIRP}{Equivalent Isotropic Radiated Power}
\newacronym{VSAT}{VSAT}{Very Small Aperture Terminal}
\newacronym{FR}{FR}{Frequency Reuse}
\newacronym{CR}{CR}{Cognitive Radios}
\newacronym{DSA}{DSA}{Dynamic Spectrum Access}
\newacronym{DL}{DL}{Deep Learning}
\newacronym{RL}{RL}{Reinforcement Learning}
\newacronym{gNB}{gNB}{Next-Generation Node B}
\newacronym{IoT}{IoT}{Internet of Things}
\newacronym{RF}{RF}{Radio Frequency}
\newacronym{MDP}{MDP}{Markov Decision Processes}
\newacronym{DRL}{DRL}{Deep Reinforcement Learning}
\newacronym{SVM}{SVM}{Support Vector Machine}
\newacronym{DQN}{DQN}{Deep Q-Networks}
\newacronym{PPO}{PPO}{Proximal Policy Optimization}
\newacronym{PSO}{PSO}{Particle Swarm Optimization}
\newacronym{MEC}{MEC}{Mobile Edge Computing}
\newacronym{GA}{GA}{Genetic Algorithms}
\newacronym{CNN}{CNN}{Convolutional Neural Networks}
\newacronym{TL}{TL}{Transfer Learning}
\newacronym{URLLC}{URLLC}{Ultra Reliable Low Latency Communication}

\usepackage{multirow}
\usepackage{array,boldline, makecell,booktabs}
\usepackage{longtable} 

    \usepackage[group-separator={,},group-minimum-digits=4]{siunitx}

\begin{document}
%
\title{Artificial Intelligence, Ambient Backscatter Communication and Non-Terrestrial Networks: A 6G Commixture}
%
%
%

\author{Muhammad Ali Jamshed,~\IEEEmembership{Senior Member,~IEEE,} Bushra Haq, Muhammad Ahmed Mohsin, Ali Nauman, Halim Yanikomeroglu, ~\IEEEmembership{Fellow,~IEEE,}
    \thanks{
        M. A. Jamshed is with the College of Science and Engineering, University	of Glasgow, UK (e-mail: muhammadali.jamshed@glasgow.ac.uk).\\
        $~~~$B. Haq is with Balochistan University of Information Technology, Engineering and Management Sciences, Pakistan (e-mail:bushra.haq@buitms.edu.pk).\\
        M. A. Mohsin is with the Department of Electrical Engineering, Stanford University, Stanford, CA, USA (email: muahmed@stanford.edu).\\
        A. Nauman is with the School of Computer Science and Engineering, Yeungnam University, Gyeongsan-si, Republic of Korea (email:anauman@ynu.ac.kr)\\
        H. Yanikomeroglu is with the Department of Systems and Computer Engineering at Carleton University, Ottawa, Canada (e-mail: halim@sce.carleton.ca)

    }
}

%
%

\markboth{Submitted to IEEE IoT Magazine}%
{Shell \MakeLowercase{\textit{et al.}}: Bare Demo of IEEEtran.cls for IEEE Journals}

\maketitle

\begin{abstract}
    The advent of Non-Terrestrial Networks (NTN) represents a compelling response to the International Mobile Telecommunications 2030 (IMT-2030) framework, enabling the delivery of advanced, seamless connectivity that supports reliable, sustainable, and resilient communication systems. Nevertheless, the integration of NTN with Terrestrial Networks (TN) necessitates considerable alterations to the existing cellular infrastructure in order to address the challenges intrinsic to NTN implementation. Additionally, Ambient Backscatter Communication (AmBC), which utilizes ambient Radio Frequency (RF) signals to transmit data to the intended recipient by altering and reflecting these signals, exhibits considerable potential for the effective integration of NTN and TN. Furthermore, AmBC is constrained by its limitations regarding power, interference, and other related factors. In contrast, the application of Artificial Intelligence (AI) within wireless networks demonstrates significant potential for predictive analytics through the use of extensive datasets. AI techniques enable the real-time optimization of network parameters, mitigating interference and power limitations in AmBC. These predictive models also enhance the adaptive integration of NTN and TN, driving significant improvements in network reliability and Energy Efficiency (EE). In this paper, we present a comprehensive examination of how the commixture of AI, AmBC, and NTN can facilitate the integration of NTN and TN. We also provide a thorough analysis indicating a marked enhancement in EE predicated on this triadic relationship.

    %

\end{abstract}


\IEEEpeerreviewmaketitle


\section{Introduction}
\label{sec:1}

\IEEEPARstart{5}{G} wireless networks have largely concentrated on enhancing transceiver designs to cope with the unpredictable nature of the wireless environment. However, as we transition to 6G, the focus is shifting towards novel concepts like \ac{NTN} and the integration of \ac{NTN} into \ac{TN}. \ac{NTN} are communication systems that utilize satellites, \ac{HAPS} and \ac{UAVs} to provide continuous connectivity to devices in remote locations such as mountainous regions, oceans, \emph{etc.}, where ground-based communication infrastructure is not available.

On one hand, \ac{NTN} plays a crucial role in offering widespread connectivity when integrated with \ac{TN}. However, there are also challenges that hinder their smooth integration, such as delays caused by increased path loss, Doppler shift, and issues with managing interference. The \ac{AmBC} has emerged as an innovative technology that enables backscatter devices to send data by harnessing existing ambient \ac{RF} signals, hence can play a crucial role in integrating \ac{NTN} and \ac{TN}. Mostly, backscatter devices are passive in nature and in some cases rely on minimal power to reflect the signal in the desired direction and are subject to various technical challenges, \emph{i.e.}, limited range, interference, \emph{etc}.

Recently, the use of \ac{AI} in wireless network shows promising results by utilizing a large number of data sets to make predictions about the network. The incorporation of \ac{AI} with \ac{NTN} and \ac{AmBC} can significantly empower these technologies to overcome their traditional limitations. Through \ac{AI}-driven signal detection, optimization, network management, and real-time decision-making, \ac{NTN} and \ac{AmBC} systems become more robust, scalable, and adaptable to dynamic environments. As the integration of \ac{AI} continues to evolve, the potential for joint utilization of \ac{NTN} and \ac{AmBC} in next-generation 6G networks will undoubtedly expand, opening the door to innovative applications and transforming the landscape of wireless communication.

In this paper, we have shown how the triad of \ac{AI}, \ac{NTN} and \ac{AmBC} can play a significant role in integrating \ac{NTN} into \ac{TN}. In Table \ref{tab:comparison}, we have provided a detailed comparison of our work with the existing literature to clearly demonstrate the contribution to knowledge. More specifically, our contributions to knowledge are summarized as follows.

\begin{itemize}
    \item We provide a detailed overview of how different variants of \ac{AI} can support the integration of \ac{NTN} into \ac{TN}.
    \item We provide a detailed overview of how different variants of \ac{AI} can overcome the challenges of \ac{AmBC}.
    \item We provide a detailed overview of how the triad of \ac{AI}, \ac{NTN} and \ac{AmBC} can support 6G.
    \item We provide a detailed overview of applications and challenges associated with the integration of \ac{AI}, \ac{NTN} and \ac{AmBC}.
    \item We present a use case demonstration of how the triad of \ac{AmBC}-\ac{UAVs}-\ac{AI}, can be used to improve the total \ac{EE} of \ac{UEs}.
\end{itemize}

\begin{table*}
    \centering
    \caption{Comparison with Existing Literature}
    \label{tab:comparison}
    \begin{tabular}{|c|c|c|c|c|c|c|c|c|c|} \hline
        Techniques                               & ~\cite{wu2022survey} & ~\cite{do2021deep} & ~\cite{mei2022intelligent} & ~\cite{leel2022reinforcement} & ~\cite{li2018uav} & ~\cite{meng2023uav} & ~\cite{das2023distributed} & ~\cite{9051982} & \textbf{Our work} \\ \hline
        AmBC                                     & \checkmark           &                    &                            &                               &                   &                     & \checkmark                 & \checkmark      & \checkmark        \\ \hline
        RIS with AmBC                            &                      &                    &                            &                               &                   &                     & \checkmark                 &                 & \checkmark        \\ \hline
        LEO-based NTN                            &                      &                    &                            & \checkmark                    &                   &                     &                            &                 & \checkmark        \\ \hline
        UAV Communications for B5G               &                      &                    &                            &                               & \checkmark        &                     &                            &                 & \checkmark        \\ \hline
        Energy Optimization                      & \checkmark           & \checkmark         & \checkmark                 &                               & \checkmark        & \checkmark          & \checkmark                 & \checkmark      & \checkmark        \\ \hline
        AI Techniques for Multi-Node Networks    &                      & \checkmark         & \checkmark                 & \checkmark                    &                   & \checkmark          & \checkmark                 &                 & \checkmark        \\ \hline
        Channel Enhancement                      &                      &                    & \checkmark                 &                               &                   & \checkmark          &                            &                 & \checkmark        \\ \hline
        Distributed Learning for 6G–IoT Networks &                      &                    &                            &                               &                   &                     & \checkmark                 &                 & \checkmark        \\ \hline
        AI + NTN + Ambient Backscatter           & \checkmark           &                    &                            &                               &                   &                     &                            &                 & \checkmark        \\ \hline
    \end{tabular}
\end{table*}

\begin{figure*}
    \centering
    \includegraphics[width=0.62\textwidth]{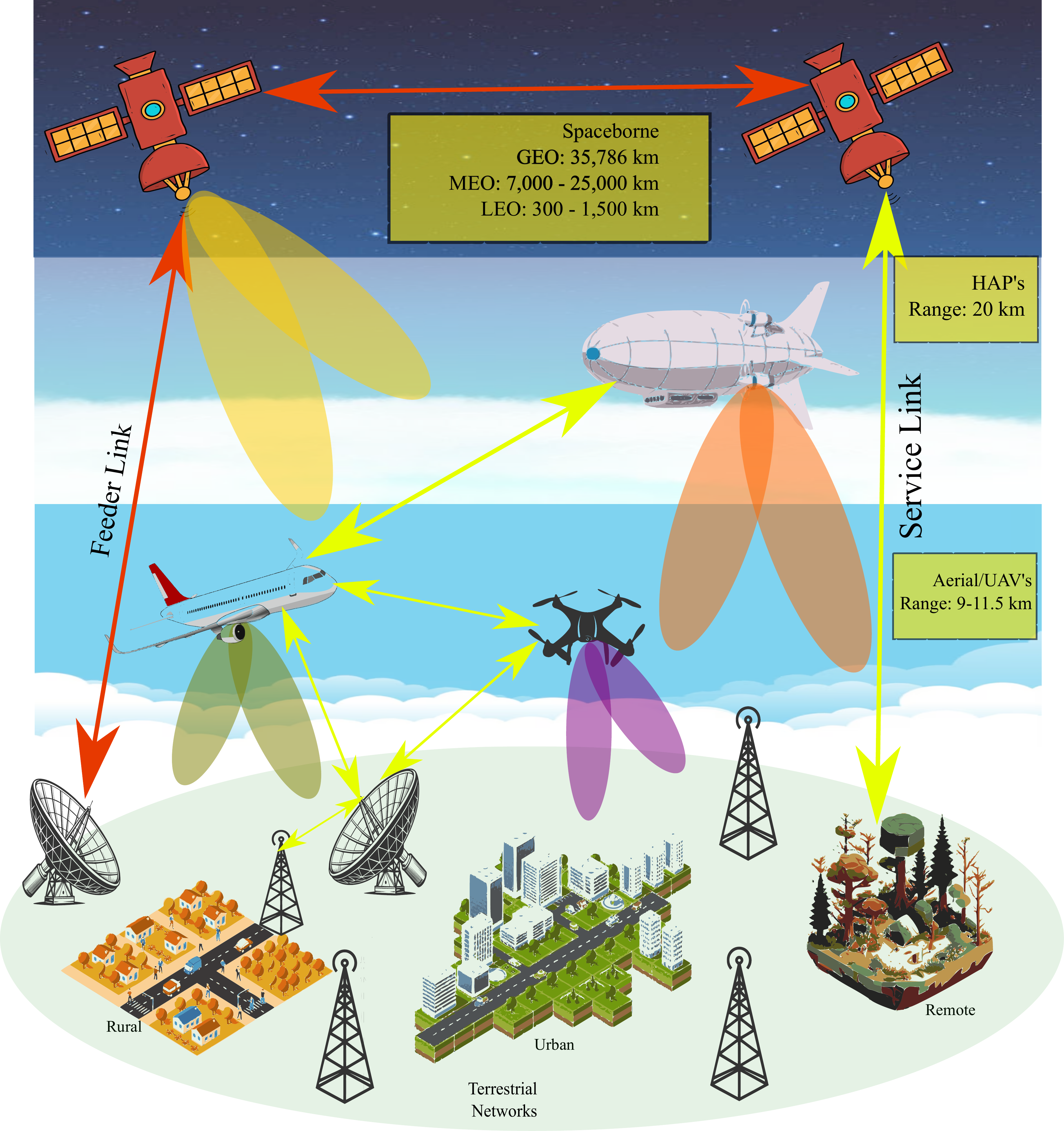}
    \caption{An illustration of \ac{3GPP}-based integrated \ac{TN} \& \ac{NTN}.}
    \label{fig:ntn}
\end{figure*}
\section{NTN \& AI}

The term \ac{NTN} is commonly linked to satellite communications, especially as the global interest in space communications has intensified in recent years. In most cases, \ac{NTN} are utilized for regions that lack or cannot easily deploy \ac{TN}. On the other hand, \ac{NTN} also include aerial platforms delivering communication services like \ac{UAVs} and \ac{HAPS} which operate within the atmosphere. These platforms fall under the broader category of \ac{NTN} which enhance the flexibility and diversity of \ac{NTN} quite considerably.


Numerous services can be executed utilizing \ac{NTN}, such as calamity recovery, search and rescue operations, and firefighting endeavors. A thorough representation of \ac{NTN} from the perspective of \ac{3GPP} is exhibited in Fig.\ref{fig:ntn}. This figure proficiently elucidates how diverse platforms, encompassing satellites, \ac{UAVs}, and \ac{HAPS}, collaborate under the \ac{NTN} framework to furnish uninterrupted communication services. On one hand, \ac{NTN} demonstrate the capability to facilitate ubiquitous connectivity when integrated with \ac{TN}. Nonetheless, there exist several obstacles associated with \ac{NTN} that restrict this integration. For instance, \ac{UAVs}, in particular, can establish network connections and transmit signals through relay systems. However, their operational duration is restricted, fluctuating from merely a few minutes to several hours due to battery limitations. Furthermore, \ac{HAPS} are deployed in the stratosphere, and can provide broader coverage areas and enhanced battery longevity, compared to \ac{UAVs}. However, the energy consumption and interference persist as the primary challenges associated with \ac{HAPS}. Additionally, satellite-based \ac{NTN} offers augmented coverage areas, in comparison to \ac{HAPS} and \ac{UAVs}, but Doppler shift, significant delays, and interference continue to present major challenges.

Wireless networks are progressively advancing toward demand-driven, autonomously reconfigurable systems. \ac{AI} has emerged as a crucial technological enabler for 6G, allowing networks to adapt dynamically to rapidly changing conditions. Much like the human brain, \ac{AI} models that leverage data can learn from experiences and replicate actions in similar contexts. This approach reduces the complexity and computational overhead compared to traditional optimization-based decision-making processes, which is particularly advantageous in \ac{NTN}, to overcome the challenges such as, energy consumption, interference management, \emph{etc}.

Also, the optimization and management of 6G-\ac{NTN} systems hold considerable promise for solutions powered by \ac{AI}. With a combination of ground-based stations, multiple satellites providing overlapping coverage, and integration of \ac{HAPS} and \ac{UAVs}, it is essential to coordinate between terrestrial and space/aerial operations to ensure uninterrupted service for users. Efficient resource management and adherence to service-level agreements are critical factors in network optimization. \ac{AI} facilitates the network's autonomous and flexible adaptation to changes in wireless channels, traffic demand variations, and mobility patterns. \ac{DL} architectures are often utilized to emulate complex algorithms, aiming to achieve a balance between performance and convergence speed. In scenarios where real-world testing is possible, \ac{RL} can be applied to explore different configurations through trial and error until optimal performance is reached.

Proactively managing network congestion and failures is essential, requiring ongoing monitoring and analysis of network performance data. \ac{AI}-based tools for detecting anomalies—such as interference or link failures—and predicting network metrics like demand, terminal trajectories, and congestion have shown great value for effective network management. These tools allow for corrective actions to be implemented before a decline in users' \ac{QoS} occurs. This is particularly critical in \ac{NTN}, where spectrum congestion is prevalent due to the increasing number of space/aerial platforms, heightening the risk of interference. 

\section{AmBC \& AI}

\ac{AmBC} is a promising wireless technology that facilitates communication between devices by reflecting and modulating ambient \ac{RF} signals, without generating new signals. It works more effectively where \ac{EE} and cost-effectiveness are critical such as low power, battery-less devices, \emph{i.e.}: \ac{IoT}. However, \ac{AmBC} encounters significant challenges including weak signal detection, limited communication range, channel interference, \emph{etc}. As a result, the full potential of \ac{AmBC}’s remains uncertain especially with the rise of \ac{AI}. \ac{AI}, with its capacity to learn from data, adapt to changing environments, and optimize system performance, has become a powerful tool to enhance \ac{AmBC} systems. This dynamic adaptability is key to unlocking the scalability, efficiency, and reliability that \ac{AmBC} needs, particularly in large-scale \ac{IoT} applications \cite{li2023physical}.

One of the critical contributions of \ac{AI} to \ac{AmBC} is signal detection and classification. \ac{AI} algorithms, such as \ac{CNN} and \ac{SVM}, can process raw backscattered signals and accurately detect weak signals amidst noise. This enhanced detection capability is particularly beneficial in noisy environments, where traditional methods struggle to distinguish between relevant communication signals and ambient interference. By improving the accuracy of signal detection, \ac{AI}-driven \ac{AmBC} systems can maintain more reliable communication links even in challenging conditions.

\ac{AI} also enables adaptive learning and optimization within \ac{AmBC} systems. \ac{RL} algorithms, such as Q-Learning and \ac{DQN}, allow \ac{AmBC} devices to learn from their communication attempts and optimize their behavior based on feedback from the environment. These algorithms dynamically adjust parameters like modulation schemes, transmission times, and power usage, optimizing the overall system performance while reducing energy consumption. This adaptive capability is essential for energy-constrained \ac{IoT} devices, where maximizing efficiency is crucial for extending operational life. \ac{AI} enables adaptive network management by analyzing vast amounts of data from backscattered signals, allowing real-time monitoring and predicting network status, making the infrastructure more reliable for 5G/6G demands.

Real-time decision-making is another area where \ac{AI} proves invaluable for \ac{AmBC}. \ac{AI} techniques, particularly \ac{RL} and optimization algorithms, can improve throughput by efficiently managing transmission resources and reducing latency through adaptive real-time adjustments to signal transmission parameters. \ac{DRL} techniques, including \ac{PPO} and \ac{DQN}, empower \ac{AmBC} systems to make intelligent, real-time decisions about communication strategies. For instance, \ac{AI} can predict the optimal transmission strategy based on current environmental conditions, such as signal strength, interference levels, or the availability of ambient \ac{RF} signals. This allows \ac{AmBC} systems to operate more efficiently by continuously adjusting their communication parameters to match the prevailing conditions.

Moreover, \ac{AI}’s role in \ac{EE} is particularly noteworthy. Optimization algorithms such as \ac{PSO} and \ac{GA} are employed to reduce the energy consumption of \ac{AmBC} devices by identifying the most efficient communication strategies. These algorithms optimize factors such as reflection angles, transmission power levels, and signal modulation schemes, allowing \ac{AmBC} systems to conserve energy while maintaining robust communication. This is especially important in large-scale \ac{IoT} deployments, where thousands of battery-less devices need to communicate over extended periods without draining energy reserves.

\ac{AI} also enhances the scalability of \ac{AmBC} networks, a critical factor for future 6G and \ac{IoT} applications. Techniques such as \ac{TL} and clustering allow \ac{AmBC} systems to scale efficiently across large, complex networks without the need for extensive retraining. \ac{TL} enables the application of knowledge gained in one communication environment to new, similar environments, reducing the time and computational resources required for optimization. This scalability is essential for implementing \ac{AmBC} in smart cities, environmental monitoring, and other large-scale deployments. Furthermore, \ac{AI} integration supports advanced applications such as backscatter-assisted relay networks, cognitive communication networks, and \ac{MEC} by optimizing communication performance and enabling efficient resource management for complex, large-scale networks.

Finally, \ac{AI} strengthens the security and robustness of \ac{AmBC} systems. By employing probabilistic models such as Bayesian Networks, \ac{AI} helps predict and adjust for uncertainties in ambient signals, ensuring more reliable communication even in unpredictable environments. These models can detect anomalies in the signal patterns, enhancing the system’s ability to prevent communication failures and detect security threats, such as eavesdropping or signal jamming. \ac{AI} also plays a key role in mitigating jamming attacks by analyzing backscattered signals in real-time and taking corrective actions to secure communication.

\section{Integrated AI, AmBC and NTN: A 6G  Vision}
The rapidly evolving world of wireless communication is set to introduce a new paradigm by integrating \ac{AI}, \ac{AmBC}, and \ac{NTN}. The triad of these technologies addresses the potential issues for future wireless networks and optimizes overall network performance. An overview of \ac{AI} algorithms for \ac{AmBC} and \ac{NTN} are presented in Table \ref{tab:ai_algorithms}.

\begin{table*}[htbp]
    \centering
    \begin{tabular}{|p{6cm}|p{9cm}|} 
        \hline
        \multicolumn{2}{|c|}{\textbf{AI Algorithms for AmBC \& NTN}}                                                                                                 \\
        \hline
        \multirow{3}{=}{\textbf{1- Machine Learning (ML) Algorithms}}    & • Supervised Learning (e.g., Support Vector Machines, Random Forest, k-Nearest Neighbors) \\
                                                                         & • Unsupervised Learning (e.g., k-Means Clustering, Principal Component Analysis)          \\
                                                                         & • Reinforcement Learning (RL) (e.g., Markov Decision Processes (MDP) )                    \\
        \hline
        \multirow{4}{=}{\textbf{2- Deep Learning Algorithms}}            & • {Convolutional Neural Networks (CNNs)}                                                  \\
                                                                         & • Recurrent Neural Networks (RNNs)                                                        \\
                                                                         & • Long Short-Term Memory (LSTM)                                                           \\
                                                                         & • {Generative Adversarial Networks (GANs)}                                                \\
        \hline
        \multirow{2}{=}{\textbf{3- Optimization Algorithms}}             & • Genetic Algorithms (GAs)                                                                \\
                                                                         & • Particle Swarm Optimization (PSO)                                                       \\
        \hline
        \multirow{2}{=}{\textbf{{4- Deep Reinforcement Learning (DRL)}}} & • Proximal Policy Optimization (PPO)                                                      \\
                                                                         & • Deep Q-Networks (DQN)                                                                   \\
        \hline
        \multicolumn{2}{|c|} {\textbf{5- Transfer Learning} }                                                                                                        \\
        \hline
        \multicolumn{2}{|c|} {\textbf{6- Bayesian Networks}}                                                                                                         \\
        \hline
    \end{tabular}
    \caption{AI Algorithms for AmBC \& NTN}
    \label{tab:ai_algorithms}
\end{table*}

\subsection{Role of AI in 6G}
Rapid development of the devices and technologies that mitigate the physical and digital world would require extravagant experiences. One of the critical elements in the 6G networks will be implementing \ac{AI} rooted solutions in managing complex dynamic environments. Automated network management using \ac{AI} technologies such as \ac{ML} and \ac{DRL} will streamline the operations of resource allocation, decision-making, and security. With the help of \ac{AI}, it is possible to manage diverse network interfaces and environments in real time, including integrated \ac{NTN} and \ac{TN} which improves service provision and network efficiency.

\ac{AI} is expected to be a determinant factor when it comes to successfully optimizing environments and operations of the 6G networks, which are most likely expected to be very complex and dynamic. Key areas to be influenced by \ac{AI} solutions include the management of networks through the use of \ac{AI} and its sub-branches such as \ac{ML}/ \ac{DRL} decision-making regarding resource management and finally enhancing security management. With \ac{AI} providing the ability of on-time functionality adaptation through optimal combination of different types of network elements including \ac{NTN} and \ac{TN}, the overall service quality and performance parameters of the network are enhanced~\cite{ndiaye2023zeroenergydevice6grealtimebackscatter}. The huge amount of data collected and analyzed by \ac{NTN} and \ac{AmBC} systems will be improved with \ac{AI}. 

\subsection{AmBC \& NTN}
\ac{AmBC} is a new means of communication that uses RF's that are already present in the area for ultra-low power communication. In 6G, it is expected that \ac{AmBC} will be an \ac{IoT} enabler and transport data to any place without a dedicated infrastructure that constitutes powered devices. It is expected that \ac{NTN} will be fundamental in widening the coverage of the 6G network to rural and other remote areas. When combined with \ac{NTN}, \ac{AmBC} can extend its reach by utilizing satellite-aided, \ac{UAVs}, and \ac{HAPS}-based platforms to reflect and transport signals over vast distances, even in challenging environments. \ac{NTN}, with their ability to provide connectivity in areas where traditional infrastructure is lacking, can complement \ac{AmBC} by providing a robust communication link in remote or underserved regions.


\subsection{Utilization of RIS with AmBC}
In essence, devices modulate the ambient signals they reflect, utilizing minimal power and occasionally even harvesting energy from the environment. The technical challenge is the reflection and signal interference. Backscattered signals are typically feeble and can effortlessly be overwhelmed by the original background signal. \ac{AI} in 6G networks can assist in filtering out noise and interference to concentrate on valuable signals that rebound off surfaces. When amalgamated with \ac{RIS}, these technologies can regulate how signals traverse the environment, directing them more efficiently to the designated receivers. This process aids in diminishing signal loss caused by obstructions and enhances both signal strength and reliability.


\subsection{Synergy of AI, AmBC, and NTN}
With the amalgamation of \ac{AI} into \ac{NTN}, the functioning of the aerial/space platforms can be performed autonomously devoid of any human intervention, facilitating intelligent resource distribution and trajectory formulation, predicated on real-time data. This will permit versatile \ac{AI}-driven \ac{NTN} for uninterrupted connectivity in diverse and extreme environments. Furthermore, with \ac{AmBC} integrated, \ac{NTN} can furnish \ac{URLLC} to \ac{IoT} devices, establishing an advanced framework for 6G applications, encompassing smart urban areas, autonomous transportation, and industrial automation. The triad of these technologies represents a significant advancement towards intelligent and sustainable 6G networks.

\section{Applications}

In 6G communications, \ac{AmBC} and \ac{NTN} powered by \ac{AI} are emerging and playing a transformative role. The \ac{AI}-based \ac{AmBC} provide low-power communication systems with unique features of \ac{EE} while \ac{NTN} offer extensive coverage beyond the terrestrial limits. The merger of these technologies; empowered by \ac{AI}, will facilitate the deployment of ubiquitous communication networks for diverse applications across smart cities, agriculture, environmental monitoring, healthcare, logistics, and more. This section explores the potential applications of \ac{AI} in \ac{AmBC} and \ac{NTN} in the context of the 6G ecosystem, drawing from case studies and real-world scenarios.

\subsection{Infrastructure Monitoring}
For the monitoring of urban infrastructures such as bridges, roads, and buildings, \ac{AI} enabled \ac{AmBC} and \ac{NTN} is a possible solution \cite{app01}. Through the integration of \ac{AI} algorithms, \ac{NTN} can process vast amounts of sensor data collected from these structures, enabling early detection of structural anomalies or degradation. Where the sensors can backscatter data about structural integrity or environmental conditions, allowing for timely maintenance and safety assessments.

\subsection{Smart Cities \& Smart Homes}
The integration of \ac{AI}-enabled \ac{AmBC} and \ac{NTN} systems in smart cities offers a seamless solution for efficient waste management, like battery-less smart bins equipped with \ac{AmBC} technology that can communicate waste levels using backscattered signals from existing ambient sources, such as \ac{Wi-Fi} or TV signals. In the smart home systems, the \ac{NTN} such as drones or \ac{HAPS} can extend the coverage where the smart home devices can communicate with a central system using \ac{AmBC} without batteries and relying on ambient signals from nearby networks or even signals transmitted by \ac{NTN}. The \ac{ML} techniques can also allow smart home devices to predict the use preferences and automate energy-efficient adjustments \cite{app03}.

\subsection{Agriculture and Farming}
Another key application of \ac{AI}-enabled \ac{AmBC} and \ac{NTN} is in precision farming where large-scale farms can deploy battery-less \ac{IoT} sensors using \ac{AmBC} technology to measure soil moisture and other environmental parameters. \ac{AI} can optimize the operation of battery-free devices that utilize ambient \ac{RF} signals for communication and can extend the operational lifetime of these devices in applications such as environmental monitoring and smart agriculture \cite{app02}. The \ac{AI}-enabled \ac{AmBC} and \ac{NTN} can also support to analyze the real-time sensor data to predict soil moisture and optimize irrigation schedules. \ac{AI}-driven \ac{NTN} to ensure continuous connectivity for precision agriculture, even in regions with limited \ac{TN} coverage, enabling farmers to make data-driven decisions, optimize resource use, and increase productivity while reducing operational costs.

\subsection{Healthcare}
In healthcare, the \ac{AI}-enabled \ac{AmBC} enables wearable medical sensors to continuously monitor patients’ vitals without requiring battery replacements, where \ac{AI} algorithms including \ac{SVM}, and Bayesian Networks play a critical role in detecting anomalies in patient data. Whereas \ac{AI}-driven \ac{NTN} ensure reliable connectivity for telemedicine services, facilitating remote consultations and medical interventions.

\subsection{Logistics}
In the logistics sector, the \ac{AI}-enabled \ac{AmBC} and \ac{NTN} can optimize cold chain logistics by embedding battery-less sensors in packaging to monitor temperature and humidity during transportation. \ac{AI}-enabled smart packaging significantly reduces spoilage and waste, particularly in transporting sensitive products like vaccines, food, and medicines. \ac{AI}-driven \ac{AmBC} utilizes surrounding \ac{RF} signals to communicate between devices without requiring a dedicated power source, enabling low-power tracking of goods. \ac{AI} algorithms analyze this data to optimize route planning, monitor shipment conditions, and predict delays. \ac{AI}-driven \ac{NTN} ensure continuous connectivity for global logistics operations, enhancing inventory management, reducing transit times, and improving overall efficiency across the supply chain.

\subsection{Environment \& Wildlife Conservation}
The \ac{AI}-enabled \ac{AmBC} and \ac{NTN} can be used to monitor the environmental parameters which can signal the onset of wildfires. The remote battery-less sensors can communicate using \ac{AmBC} technology, relying on ambient \ac{RF} signals, where \ac{NTN} can provide continuous connectivity, ensuring uninterrupted data collection and transmission. \ac{DRL} models can process this data to detect early signs of a fire, while \ac{MDP} predict the spread of the fire \cite{app04}. In wildlife conservation, the \ac{AI}-enabled \ac{AmBC} can be attached to the animals, transmitting data on their movements and environmental conditions supported by \ac{NTN}. The \ac{AI} models, including \ac{RL} and \ac{TL}, can analyze animal movement patterns and predict migration paths to monitor animal behavior and respond to potential threats like the destruction of habitat.
\subsection{Smart Traffic Communication}
The \ac{AI} can also facilitate the communication between the vehicles and their surroundings using the \ac{AmBC} and \ac{NTN} which can improve traffic management, enhance safety features, and enable smart transportation systems. \ac{AI}-driven \ac{AmBC} uses ambient \ac{RF} signals for low-power data transmission between vehicles and surrounding infrastructure, reducing energy consumption. \ac{AI}-driven \ac{NTN} ensure continuous connectivity in urban and remote areas, supporting the scalability of smart transportation systems and improving overall traffic management efficiency.


\section{Challenges and Research Directions}

\subsection{Challenges}
The integration of \ac{AI} with \ac{AmBC} and \ac{NTN} presents several challenges, especially when these systems are operated within the emerging 6G framework.
\begin{enumerate}
    \item \textbf{Signal Detection and Estimation:}  One of the major tasks within these systems is signal detection and estimation in the presence of noise and interference. While \ac{AI} can improve signal processing, accurately detecting and estimating backscattered signals in complex environments remains difficult. Existing algorithms may fall short when faced with multipath propagation or unpredictable \ac{RF} conditions, particularly in large-scale \ac{IoT} deployments.
    \item \textbf{\ac{EE} and Power Constraints:} While \ac{AmBC} relies on low-energy harvesting from ambient signals, \ac{NTN} typically involve satellites or \ac{HAPS} with strict power limitations. Integrating \ac{AI} for real-time optimization and decision-making requires balancing these power constraints, especially as \ac{AI} algorithms can be computationally intensive, leading to increased energy consumption in systems designed for low-power operations.
    \item \textbf{Latency and Real-Time Processing:} \ac{AmBC} and \ac{NTN} encounter latency issues due to long propagation delays (especially in \ac{NTN}) and low data rates (in \ac{AmBC}). \ac{AI}-based solutions must address these latency challenges, as real-time decision-making in 6G networks becomes critical for applications like autonomous vehicles and remote healthcare, requiring optimized \ac{AI} models that can operate effectively within these delayed communication environments.
    \item \textbf{Scalability and Connectivity:} As 6G networks aim to support massive \ac{IoT} connectivity, scaling \ac{AmBC} and \ac{NTN} presents major challenges. The integration of \ac{AI} for managing millions of connected devices in \ac{NTN}, alongside the low-power, low-data-rate \ac{AmBC} nodes, requires innovative \ac{AI}-driven network management algorithms capable of optimizing resource allocation, minimizing congestion, and ensuring seamless connectivity across both terrestrial and non-terrestrial links.
    \item \textbf{Security and Privacy:} In \ac{AI}-empowered \ac{NTN} and \ac{AmBC} security concerns, such as jamming and ensuring the privacy of communications, are critical, particularly in applications with sensitive data. \ac{AI} algorithms used for data processing, authentication, and anomaly detection must be robust enough to handle vulnerabilities from both terrestrial and space-borne threats. Securing communication in resource-constrained \ac{AmBC} systems while maintaining the integrity of \ac{NTN} transmissions (especially with satellite links) is a key challenge that requires advanced \ac{AI}-based cryptographic methods and threat detection systems.
    \item \textbf{Heterogeneous Network Coordination:} Both \ac{AmBC} and \ac{NTN} operate within highly heterogeneous environments in 6G, with diverse devices, frequencies, and communication protocols. \ac{AI}-based techniques must be capable of coordinating these networks seamlessly, ensuring smooth transitions between terrestrial, aerial, and space-based nodes while maintaining consistent \ac{QoS} across the network.
\end{enumerate}
\begin{figure*}[!t]
    \centering
    \begin{subfigure}[t]{0.45\textwidth}
        \centering
        \includegraphics[width=\columnwidth]{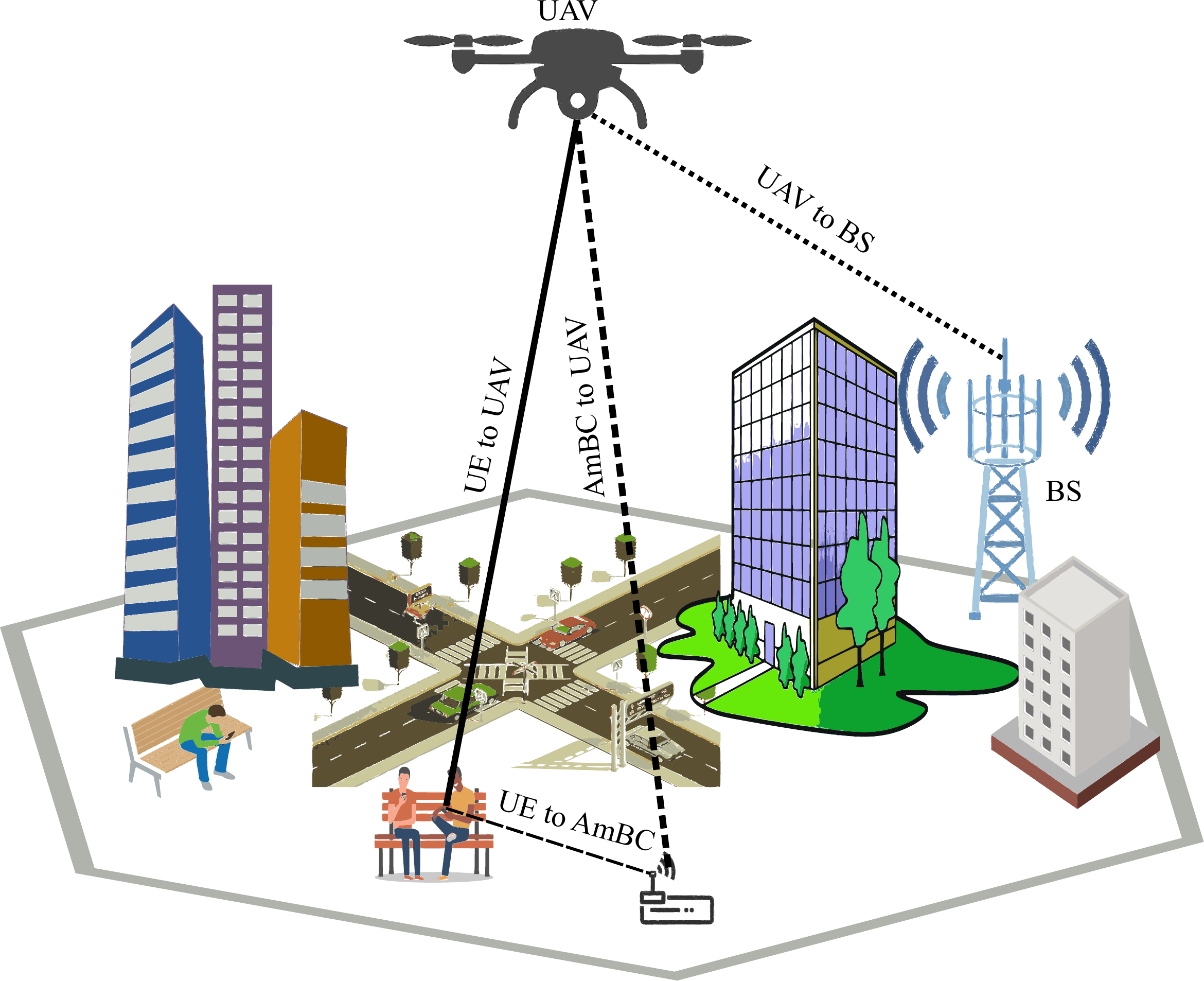}
        \caption{}
        \label{fig11}
    \end{subfigure}
    \vfill
    \begin{subfigure}[t]{0.45\textwidth}
        \centering
        \includegraphics[width=\columnwidth]{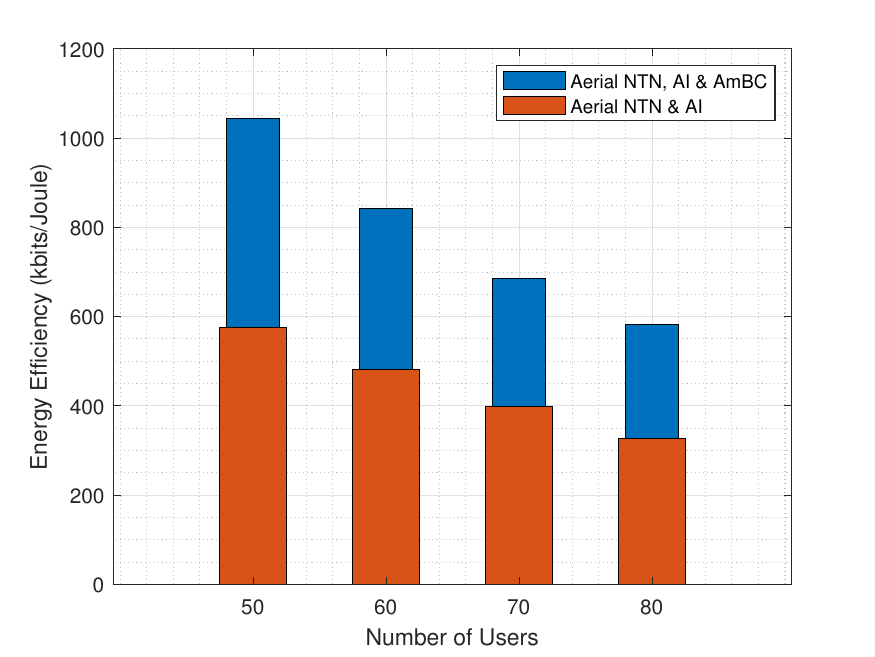}
        \caption{}
        \label{fig22}
    \end{subfigure}
    \hfill
    \begin{subfigure}[t]{0.45\textwidth}
        \centering
        \includegraphics[width=\columnwidth]{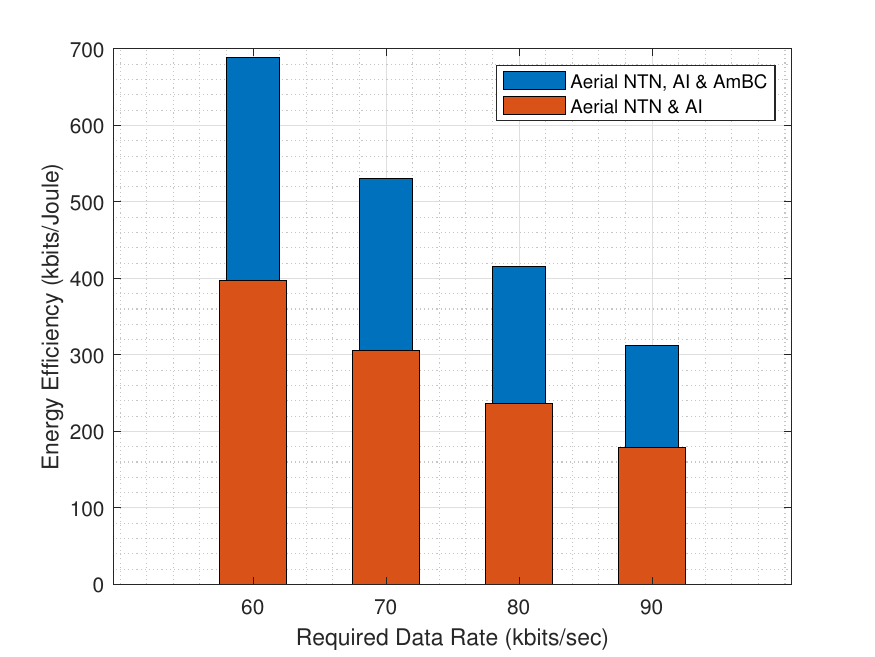}
        \caption{}
        \label{fig33}
    \end{subfigure}
    \caption{(a) System model of the proposed triad of \ac{AmBC}-\ac{UAVs}-\ac{AI} aided communication, (b) \ac{EE} versus the varying number of users for a fixed transmitted data size, (c)\ac{EE} versus the transmitted data size for a fixed number of UEs.}
\end{figure*}

\subsection{Research Directions}
Research directions in applying \ac{AI} to \ac{AmBC} and \ac{NTN} within the 6G ecosystem must focus on several key areas. Advanced \ac{AI} algorithms tailored for signal detection, channel estimation, and interference management are crucial for enhancing system performance in complex environments. Exploring policy-based and value-based algorithms will help address challenges related to security, jamming, and throughput, providing innovative solutions for real-time decision-making in dynamic conditions. Additionally, research into hybrid communication models that combine backscatter with traditional \ac{RF} communication can improve system reliability and scalability, particularly in large-scale \ac{IoT} applications. Real-time data processing powered by \ac{AI} will further enhance the responsiveness of \ac{AmBC} and \ac{NTN} systems while integrating \ac{RIS} with backscatter communications can improve signal quality and coverage. Application-specific solutions tailored to domains like smart cities, healthcare, and environmental monitoring will help address unique challenges and leverage \ac{AI}'s strengths. Lastly, establishing performance evaluation frameworks will be essential to assess the effectiveness of \ac{AI} algorithms in these networks, guiding future advancements in \ac{AI}-empowered \ac{AmBC} and \ac{NTN} for 6G applications.

\section{Design and Results}

In this section, the efficacy of the triad comprising \ac{AI}, \ac{UAVs}, and \ac{AmBC} has been illustrated through a comprehensive simulation analysis. As depicted in Fig. \ref{fig11}, we have posited a densely populated urban uplink scenario wherein users are incapacitated in their attempts to communicate with the \ac{BS}, thereby necessitating the joint utilization of \ac{UAVs} and \ac{AmBC} to facilitate a connection with the \ac{BS}. The \ac{EE} of \ac{UEs} have been optimized by leveraging \ac{NOMA}, \ac{AI}, \ac{AmBC}, and a UAV. As delineated in \cite{jamshed2021unsupervised}, the initial step involves capitalizing on the advantages of \ac{NOMA} through the execution of user grouping and subcarrier allocation utilizing a collective approach that encompasses k-means clustering, F-test analysis, and the elbow method. Subsequently, power allocation is conducted employing iterative methodologies to enhance the \ac{EE} of \ac{UEs}. We have instituted a coverage radius of 300 meters, fixed the number of subcarriers at 128, established a bandwidth of 1 MHz, the maximum uplink transmit power is fixed at 0.2 Watts, maintained the UAV altitude at 100 meters, and adopted the path loss model articulated in \cite{jamshed2021unsupervised}.

In Fig. \ref{fig22}, an analysis has been conducted on the efficacy of the proposed triad of \ac{AmBC}-UAV-\ac{AI} in relation to the fluctuation in the quantity of \ac{UEs} present within a cellular area, while the data transmitted by each UE remains constant at 60 kbits, and the circuit power is maintained at 5 dBm. As illustrated in Fig. \ref{fig22}, the augmentation in the number of devices correlates with a decrease in the \ac{EE} of the system, culminating in the formation of an exponential curve. When comparing with a reliance solely on \ac{AI} and UAV, the triad of \ac{AmBC}-UAV-\ac{AI} yields a notable improvement in \ac{EE}, showcasing the superiority of the triad.

In Fig. \ref{fig33}, a comprehensive examination has been performed regarding the effectiveness of the proposed triad of \ac{AmBC}-UAV-\ac{AI} in relation to the variations in the data transmitted by each UE located within a cellular environment, while the number of \ac{UEs} remains invariant at 70, and the circuit power is sustained at 5 dBm. Generally, an increase in the volume of data transmitted, given a constant number of \ac{UEs}, results in a decline in the overall \ac{EE}. When compared with the reliance on \ac{AI} and UAV, the triad of \ac{AmBC}-UAV-\ac{AI} consistently enhances the \ac{EE} across all volumes of transmitted data. Approximately, the performance differential maintains consistency throughout in Fig.~\ref{fig33}, which can be attributed to the constant circuit power value. It is observed that circuit power exerts a predominant influence in attaining an elevated value of \ac{EE}. Notwithstanding, the triad of \ac{AmBC}-UAV-\ac{AI} possesses the capability to augment the \ac{EE} for a substantial number of \ac{UEs}. 

\section{Conclusion}

This article proposes to utilize the triad of AI, AmBC, and NTN to support the full integration of NTN into TN. Firstly, we provided a detailed overview of the challenges associated with NTN and how we leverage AI to enable the successful integration of NTN into TN. Secondly, we provided an overview of AmBC technology and the challenges associated with the use of AmBC in TN. We also discuss how AI can support the full adoption of AmBC in future TN. Thirdly, we demonstrate how the AI, AmBC, and NTN triad can facilitate the amalgamation of NTN into TN. Fourthly, we provided some applications and challenges associated with this triad. Finally, we provide a case study to showcase the superiority of this triad compared to relying on AI and NTN.


\begin{IEEEbiographynophoto}{Muhammad Ali Jamshed} (Senior Member, IEEE) received a Ph.D. degree from the University of Surrey, Guildford, U.K, in 2021. He is with University of Glasgow, since 2021. He is endorsed by Royal Academy of Engineering under exceptional talent category and was nominated for Departmental Prize for Excellence in Research in 2019 and 2020 at the University of Surrey. He is a Fellow of Royal Society of Arts, a Fellow of Higher Education Academy UK, a Member of the EURASIP Academy, and an Editor of IEEE Wireless Communication Letter and an Associate Editor of IEEE Sensor Journal, IEEE IoT Magazine, and IEEE Communication Standard Magazine. His research interests are energy efficient IoT networks, AI for wireless communication, EMF exposure measurements, and backscatter communications.
\end{IEEEbiographynophoto}

\vspace{-10mm}

\begin{IEEEbiographynophoto}{Bushra Haq} is an Assistant Professor at the Balochistan University of Information Technology, Engineering, and Management Sciences (BUITEMS), Quetta, Pakistan. She earned her M.S. degree from BUITEMS, where she is currently pursuing her Ph.D. and has been an active member of the academic and research community since 2014. Her research interests lie in the areas of machine learning, deep learning, data science, and data mining. She has made significant contributions to these fields, which are reflected in her publications that have collectively garnered over 50 citations. Through her work, she continues to advance knowledge and innovation, focusing on developing solutions for real-world challenges using advanced computational techniques. In addition to her research endeavors, Bushra Haq is committed to academic excellence and mentoring, playing a pivotal role in shaping future scholars and professionals at BUITEMS. Her dedication to both research and teaching underscores her multifaceted contributions to the academic community.

\end{IEEEbiographynophoto}

\begin{IEEEbiographynophoto}{Muhammad Ahmed Mohsin}received the B.E.degree in electrical engineering from National University of Sciences andTechnology (NUST), Pakistan. He is currently pursuing the Ph.D. degree in  electrical engineering from Stanford University, USA. His primary focus of research lies in next-generation wireless communications, deep learning (DL), and reinforcement learning (RL).

\end{IEEEbiographynophoto}

\begin{IEEEbiographynophoto}{Ali Nauman} received the M.Sc. degree in wireless communications from the Institute of Space Technology, Islamabad, Pakistan, in 2016, and the Ph.D. degree in information and communication engineering from Yeungnam University, Gyeongsan, Republic of Korea, in 2022. He is currently working as an Assistant Professor with the Department of Computer Science, Yeungnam University. He has contributed to five patents and authored/co authored five book chapters and more than 75 technical articles in leading journals and peer-reviewed conferences. Dr. Nauman has also edited two books and serves as an editor and a reviewer of highly reputed journals and conferences.

\end{IEEEbiographynophoto}

\begin{IEEEbiographynophoto}{Halim Yanikomeroglu} (Fellow, IEEE) is a Chancellor’s Professor in the Department of Systems and Computer Engineering at Carleton University, Canada; he is also the Director of Carleton-NTN (Non-Terrestrial Networks) Lab. He is a Fellow of IEEE, Engineering Institute of Canada (EIC), Canadian Academy of Engineering (CAE), and Asia-Pacific Artificial Intelligence Association (AAIA). He is a Distinguished Speaker for IEEE Communications Society and IEEE Vehicular Technology Society, and an Expert Panelist of the Council of Canadian Academies (CCA|CAC). Dr. Yanikomeroglu has coauthored 650+ papers including 320+ published in 31 different IEEE journals; he also has 41 granted patents. He has supervised or hosted at Carleton 165+ postgraduate researchers; several of his former team members have become professors in Canada, US, UK, and around the world. He has given around 110 invited seminars, keynotes, panel talks, and tutorials in the last five years. He has served as the Steering Committee Chair, General Chair, and Technical Program Chair of a high number of major international IEEE conferences, as well as in the editorial boards of several IEEE periodicals. Dr. Yanikomeroglu received many awards for his research, teaching, and service. He holds a PhD degree in electrical and computer engineering from University of Toronto.

\end{IEEEbiographynophoto}

\end{document}